\documentclass[conference]{IEEEtran}
\IEEEoverridecommandlockouts
\usepackage{cite}
\usepackage{amsmath,amssymb,amsfonts}
\usepackage{amsthm}
\usepackage{algorithmic}
\usepackage{graphicx}
\usepackage{textcomp}
\usepackage{xcolor}
\def\BibTeX{{\rm B\kern-.05em{\sc i\kern-.025em b}\kern-.08em
    T\kern-.1667em\lower.7ex\hbox{E}\kern-.125emX}}

\usepackage{enumitem}
\usepackage{multirow}
\usepackage{subfigure}
\usepackage[linesnumbered,ruled,vlined]{algorithm2e}
\SetEndCharOfAlgoLine{;}
\begin{document}

\title{A User-to-User Resource Reselling Game in Open RAN with Buffer Rollover}

\author{
    \IEEEauthorblockN{Ruide Cao$^{1,2}$, Marie Siew$^{*1}$, David Yau$^{*1}$}
    \IEEEauthorblockA{$^1$Singapore University of Technology and Design}
    \IEEEauthorblockA{$^2$University of California, Riverside}
}

\maketitle

\renewcommand{\thefootnote}{\fnsymbol{footnote}}
\footnotetext{$^*$Corresponding authors: Marie Siew (marie\_siew@sutd.edu.sg) and David Yau (david\_yau@sutd.edu.sg). This research is supported by the National Research Foundation, Singapore and Infocomm Media Development Authority under its Future Communications Research \& Development Programme, and by the MOE START Faculty Early Career Award supported by SUTD and the Singapore Ministry of Education, and in part by SUTD Kickstarter Initiative (SKI) grant with No. SKI 2021\_04\_07.}

\begin{abstract}
The development of the Open RAN (O-RAN) framework helps enable network slicing through its virtualization, interoperability, and flexibility. To improve spectral efficiency and better meet users' dynamic and heterogeneous service demands, O-RAN's flexibility further presents an opportunity for resource reselling of unused physical resource blocks (PRBs) across users. In this work, we propose a novel game-based user-to-user PRB reselling model in the O-RAN setting, which models the carryover of unmet demand across time slots, along with how users' internal buffer states relate to any PRBs purchased. We formulate the interplay between the users as a strategic game, with each participant aiming to maximize their own payoffs, and we prove the existence and uniqueness of the Nash equilibrium (NE) in the game. We furthermore propose an iterative bidding mechanism that converges to this NE. Extensive simulations show that our best approach reduces data loss by 30.5\% and spectrum resource wastage by 50.7\% while significantly improving social welfare, compared to its absence.
\end{abstract}


\section{Introduction}

The Open Radio Access Network (Open RAN or O-RAN) framework represents a paradigm shift in the architecture and deployment of mobile networks, with the aim of fostering greater openness, flexibility, and intelligence beyond the traditionally closed and proprietary RAN ecosystem \cite{polese2023understanding}. It allows a mobile network operator (MNO) to integrate hardware and software components from multiple vendors. Its virtualization and interoperability features also provide a ready foundation for flexible {\em slicing} of network resources. The slicing multiplexes a number of virtual partitions onto the network's shared physical resources, each slice tailored to meet specific service requirements. For example, one slice may give a high share of communication bandwidth to support bulk data transfer. Another slice may be given a low bandwidth share for intermittent communication, but a high buffer share to absorb the burstiness of instantaneous demand.

To improve network efficiency, cost effectiveness, and satisfaction of diverse service demands of heterogeneous users, optimizing the allocation of radio resources - such as physical resource blocks (PRBs or physical RBs) - is a key research challenge \cite{polese2023understanding, filali2024open,motalleb2022resource}. For instance, prior work~\cite{filali2024open} has proposed a two-timescale PRB resource allocation approach, from MNOs to Mobile Virtual Network Operators (MVNOs), and then from MVNOs to end users. A two-stage algorithm that optimizes PRB and power allocation has also been investigated~\cite{motalleb2022resource}.

At the same time, mobile network demand exhibits continual dynamism due to a multitude of factors \cite{cao2023poster, cao2024adaptive}, including changing user behavior throughout the day (e.g., peak usage during commute times or lunch breaks) and occurrences of planned or unplanned events (such as concerts, sports games, or emergencies). These dynamics may lead to surges in communication demand or evolving traffic patterns driven by new applications and their usage (e.g., the increasing popularity of video streaming or real-time gaming). They manifest in terms of required bandwidth, latency, and the number of connected devices attempting to access the network at any given time. In turn, the \textit{need for PRB radio resources changes across users and across time}. It is thus necessary to provide flexible and adaptive network management to ensure consistent performance and quality of service (QoS) for all the users \cite{polese2023understanding}. 

The heterogeneous, hence possibly complementary, nature of network demand, coupled with O-RAN's inherent flexibility, presents an opportunity for market-driven resource resale mechanisms. Inspired by existing work on resource reselling or trading that concerns mobile data, edge computing, and cloud computing \cite{siew2023towards,tang2016double}, in this paper we investigate the resale of PRB resources in an O-RAN setting. When certain users possess surplus PRBs, they may be incentivized to sell them to other users experiencing a shortfall, due to surging demand. Such trading can improve the spectral efficiency of the network by reducing the wastage of PRBs while meeting the QoS requirements of heterogeneous users. At the same time, O-RAN's highly virtualized and programmable architecture facilitates online transfers of the PRBs in question across logically isolated network slices.

To the best of our knowledge, resource resale has not yet been well solved in an O-RAN resource optimization context, unlike other network resource markets. Prior work~\cite{chen2018game} has used a game to model EVs buying (via charging) and selling (via discharging) electricity over an electrical grid. They investigate a user-to-grid resale paradigm facilitated by the grid or an MNO. In contrast, other works has considered a user-to-user resale paradigm \cite{yu2017mobile,tang2016double,siew2020dynamic, siew2023towards}, e.g., a mobile data trading market via a behavioral economics lens~\cite{yu2017mobile}; a double-sided bidding mechanism to facilitate resource trading among mobile devices in a mobile cloud~\cite{tang2016double}; optimized resource quota sharing in mobile edge computing~\cite{siew2020dynamic}; and a Stackelberg game model for reservation-plan users to resell their unused resource quota to on-demand users in edge computing~\cite{siew2023towards}. These works do not consider rollover effects across time, focusing rather on immediate resource needs within each time slot only, without addressing important carryover of pent-up demand or accumulating unused resources over time. An effort to explore this rollover effect concerns the economic viability of a mobile data trading market~\cite{wang2019economic}. Targeting a different application domain than ours, they do not consider salient features of the O-RAN setting, such as online finite buffering of user data and how it interacts with the utilization of PRBs according to channel conditions.

\begin{figure*}[tb]
    \centering
    \begin{minipage}{0.4\textwidth}
    \includegraphics[scale=0.2]{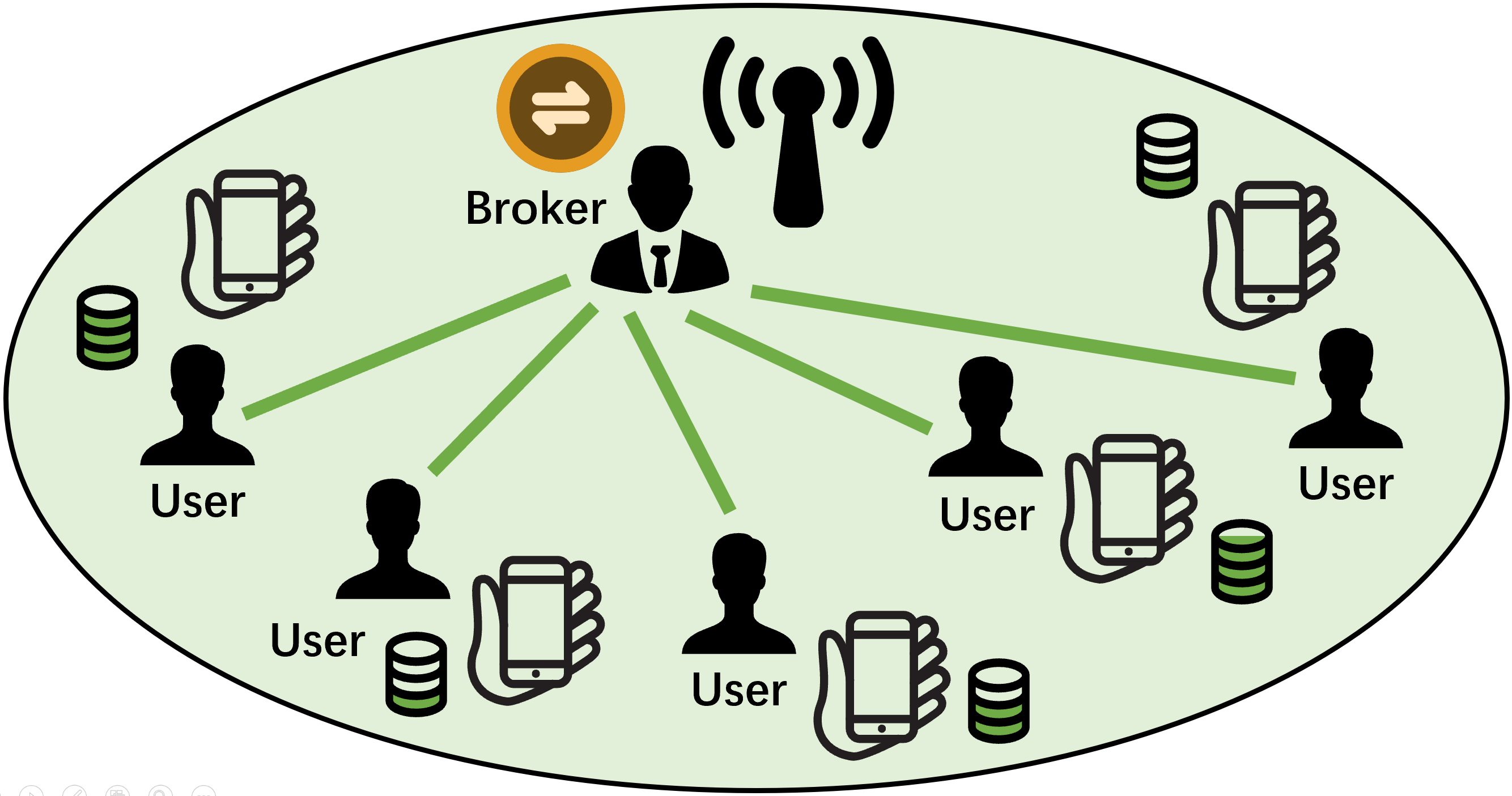}
    \caption{
        \textbf{System Model}. Users buy and sell RBs from each other through a public interest broker at the base station.
    }
    \end{minipage}
    \hspace{10pt}
    \begin{minipage}{0.52\textwidth}
    \includegraphics[scale=0.209]{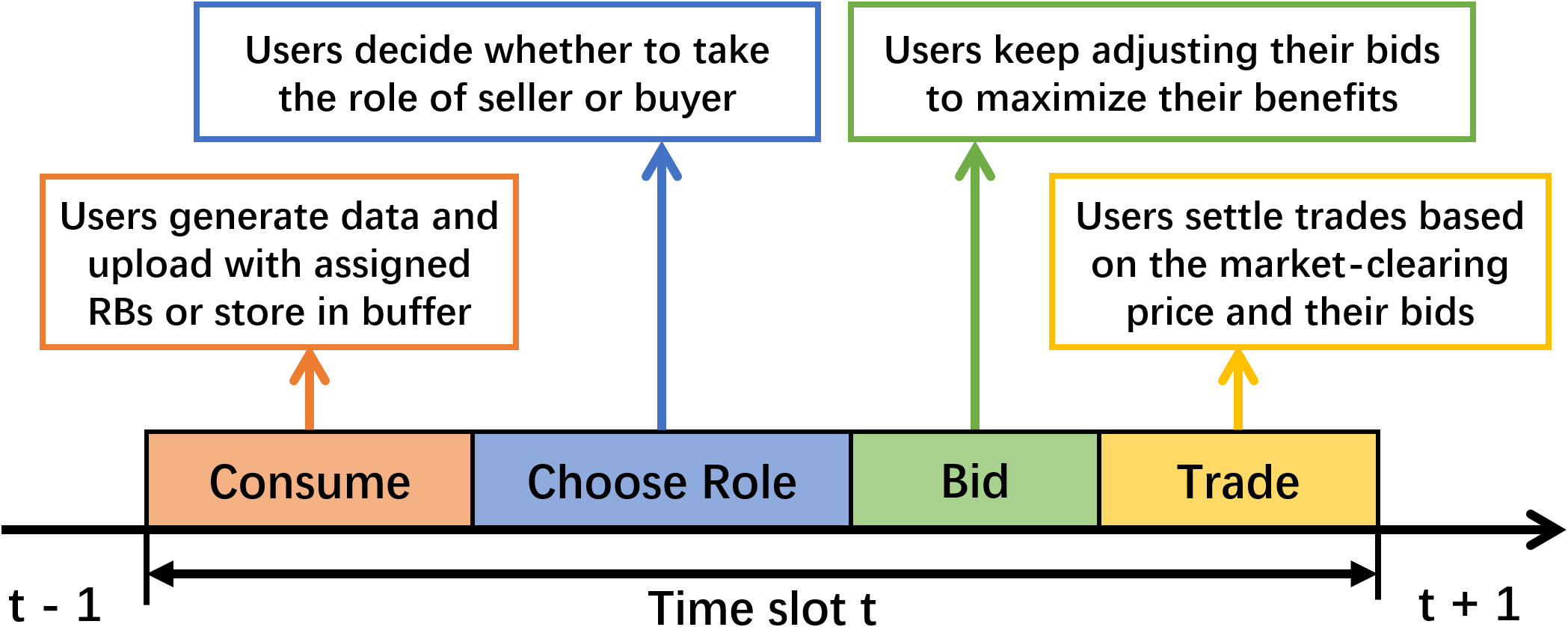}
    \caption{
        \textbf{Time Slot}. Each user maintains the role within the time slot after the \textit{Choose Role} phase, and can switch roles across different time slots.
    }
    \end{minipage}
    \vspace{-0.2cm}
\end{figure*}

Our contributions are as follows:
\begin{itemize}
    \item We propose a novel user-to-user PRB resale model in the O-RAN setting, which considers the carryover of unmet or surplus demand across time slots, under evolving per-user buffer states influenced in turn by work arrival and service through online utilization of available (including possibly purchased) PRBs.
    \item We formulate the interaction among the users as a strategic game and propose an effective 5G resource resale mechanism accordingly. In this mechanism, mobile users can dynamically trade their RB quotas with each other, depending on their current buffer availability and concomitant predicted future losses. Our game-based design enables the market to improve the overall social welfare, despite selfish users who consider their own benefits only. We prove the existence and uniqueness of the game's Nash equilibrium (NE).
    \item We present experiments that illustrate and validate the performance of the proposed mechanism. The results show that our solution can significantly reduce data loss, reduce resource wastage, and increase social welfare, compared with baselines such as no trading and random-role trading.
\end{itemize}

The rest of this paper is organized as follows. We introduce our system model in Sec. \ref{chap_system}. We present the formulation of our game model and the proposed resale mechanism in Sec. \ref{chap_reselling}. We present diverse experimental results in Sec. \ref{chap_evaluation}. Sec. \ref{chap_conclusion} concludes the paper.

\section{System Model}
\label{chap_system}

We consider multiple mobile users covered by a single cellular base station, as shown in Fig. 1. Every user faces the arrival of data transmission tasks that require the service of available PRBs. For instance, tasks involving high-throughput data transfer, like 4K video streaming and large file downloads, or low-data-rate interactive tasks, like messaging or online transactions. Each user has an allocation of buffers to store their data temporarily pending communication. Spectrum resources are assigned to them in the unit of RB. Each user has their own RB quota, which is negotiated in advance with the MNO. In addition, they can trade these quotas among each other according to their real-time needs.

We define the user set $\mathcal{U}=\{u_1, u_2, ..., u_n\}$. Time is divided into multiple time slots $\mathcal{T} = \{t_1, t_2, ..., t_m\}$. Time slot $t_i$ starts from $t_i^s$ and ends at $t_i^e$. The length of each slot is consistent, \textit{i.e.}, $\forall t_i \in \mathcal{T}, t_i^e-t_i^s = T$. The 2D coordinates of user $u_i$ at the time $t$ are given by $\textbf{u}_i(t) = [x_{u_i}(t), y_{u_i}(t)]^{tr}$, while the 2D coordinates of the base station are fixed at $\textbf{n} = [x_n, y_n]^{tr}$. We denote the station's height as $H$. Hence the distance between user $u_i$ and the base station at time $t$ can be written as
\begin{equation*}
d_i(t)=\sqrt{\left \| \mathbf{u}_i(t)-\mathbf{n} \right \|_2^2 +H^2}.
\end{equation*}
Following the Friis transmission equation, the distance can be used to determine the power of the signal received by the base station from the user:
\begin{equation*}
P_i^{rx}(t) = P^{tx} * (\frac{\lambda}{4 \pi d_i(t)})^2,
\end{equation*}
where wavelength $\lambda = c / f_c$ and $c$ is the speed of light, $f_c$ is the signal frequency, $P^{tx}$ is the transmission power on the user side.
We assume that there are unit gains for both the transmit and receive antennas, following which the transmission efficiency factor of $u_i$ within time slot $t_j$ can be calculated by 
\begin{equation*}
f_i^j = \int_{t_j^s}^{t_j^e}W * log_2(1+\frac{P_i^{rx}(t)}{N}) dt,
\end{equation*}
where $W$ is the bandwidth of one RB, and $N$ is the white Gaussian noise of the environment. With the 
user $u_i$ assigned $q_i^t \in [0, Q_{max}]$ RBs in the $t$th time slot, its corresponding \textit{uplink throughput} within the time slot would be $r_i^t = f_i^t * q_i^t$.

Each user $u_i$ has a certain amount of buffer $B_i$ for storing the data to be uploaded in a FIFO manner. We denote user $u_i$'s occupied buffer level at the $t$th time slot by $o_i^t \in [0, B_i]$ and its available buffer by $e_i^t \in [0, B_i]$. $\forall t_j \in \mathcal{T}, e_i^j + o_i^j = B_i$. These buffers can help users avoid data loss, since they may not have sufficient RBs to immediately offload all the data to be offloaded. Users may also strategically buffer more data and sell their RB quota for monetary gain.

In each time slot, new data may arrive for user $u_i$, $d_i^t \in [0, D_{max}]$. We have $c_i^t = d_i^t-r_i^t=d_i^t-f_i^t * q_i^t$, where $c_i^t \in [-f_i^tQ_{max}, D_{max}]$. $c_i^t$ equals the newly arrived data to be uploaded minus the uplink throughput, a function of the number of RBs the user currently has. A positive $c_i^t$ means that $u_i$ has generated more data to be uploaded than $u_i$'s originally assigned RBs can transmit in the time slot, while a negative $c_i^t$ indicates $u_i$'s RB quota more than covers all the data that $u_i$ has generated. On top of the original quota, depending on a user's resource needs, a user can buy $(a_i^t \geq 0)$ or sell $(a_i^t \leq 0)$ RBs in the $t$th time slot. Our reselling mechanism generates a transaction of $a_i^t$ RBs for $u_i$, resulting in a change in uplink throughput of $f_i^t a_i^t$. For the next slot, the occupied buffer will be updated according to $o_i^{t+1} = max(0, min(B_i, o_i^t+c_i^t-f_i^ta_i^t))$. \textbf{Data loss} occurs if the net change in the occupied buffer level exceeds the available empty space, represented by $l_i^t = (c_i^t - f_i^ta_i^t - e_i^t)^+$. Similarly, \textbf{RB wastage} occurs if the service capacity from RBs exceeds the total data present in the buffer: $w_i^t = (f_i^ta_i^t - c_i^t - o_i^t)^+$ (i.e., RBs not used). Users aim to minimize their data loss, \textit{i.e.}, avoid buffer overflow.

We measure user satisfaction with the transmission process using a utility function $U_i^t(\cdot)$ that is required to satisfy three properties: (1) Data loss dissatisfies the users, so the utility function strictly decreases with respect to the loss $l_i^t$. (2) To characterize diminishing marginal returns, the function is concave. (3) The user has a ``smooth'' reaction to the loss, so that the function is continuously differentiable.

\section{Reselling Mechanism}
\label{chap_reselling}

In this section, we first formulate the resale process, which considers buffer allocation and data loss, as a multi-round user-to-user game (Sec. III A). We then prove the existence and uniqueness of NE in the game (Sec. III B). Lastly, we present an iterative bidding algorithm to reach the NE (Sec. III C).

\subsection{Game Formulation}

Based on the system model, we formulate a reselling game across all the users, where each of them seeks to maximize their own payoffs by buying or selling their RB quota. Fig. 2 shows the 4 phases of a time slot: consume, choose role, bid, trade. Within the $t$th time slot, all the buyers form a buyer set $\mathcal{B}^t$, while all the sellers form a seller set $\mathcal{S}^t$. We denote user $u_i$'s bid as $b_i^t$ at the $t$th time slot, where $p^t$ is the market price. 
\begin{align}
\forall u_i \in \mathcal{B}^t,\ b_i^t &= p^t*a_i^t,\label{buyer_bid}\\
\forall u_j \in \mathcal{S}^t,\ b_j^t &= p^t*(a_j^t+q_j^t). \label{seller_bid}
\end{align}
Note that for any user $u_i$, $-q_i^t \le a_i^t \le \sum_{u_j\in \mathcal{S}^t}q_j^t$ since a seller's maximum sale is the whole originally assigned quota, while a buyer's maximum buy is the sum of all the sellers' quotas in the market. Sellers have non-positive $a_i^t$ values while buyers have non-negative ones, and all users have non-negative bids. Based on $u_i$'s utility function $U_i^t(l_i^t(a_i^t))$ (hereafter abbreviated as $U_i^t(a_i^t)$), the user's payoff function is 
\begin{equation*}
V_i^t(a_i^t) = U_i^t(a_i^t) - p^t * a_i^t.
\end{equation*}

\textit{Definition 1}: A bid case $\textbf{b}^t$ is a Nash Equilibrium if no user can improve its payoff by unilaterally changing the bid, \textit{i.e.},
\begin{equation*}
V_i^t(b_i^t, \textbf{b}_{-i}^t) = \max_{\hat{b}_i}V_i^t(\hat{b}_i, \textbf{b}_{-i}), \ \forall u_i \in \mathcal{U},
\end{equation*}
where $\textbf{b}^t$ denotes the bid case for all users at the $t$th time slot, and $\textbf{b}_{-i}^t$ denotes the bid case for all the users other than $u_i$.

\subsection{The Existence and Uniqueness of Nash Equilibrium}

We first present a lemma regarding the existence of the NE in this game. The game occurs during the Bid phase (Fig. 2). For the NE to exist, we assume that the users do not switch roles during the bid phase.

\noindent \textit{Lemma 1}: When there is only one seller, i.e., $|S^t|=1$, there is no Nash equilibrium.

We refer the reader to the appendix for the proof of this lemma. With Lemma 1, we then present a theorem on the existence and uniqueness of the NE, followed by its proof.

\textit{Theorem 1}: When there is at least one buyer and two sellers\footnote{If there is no buyer or seller,  resale cannot take place. If there exists one or more buyers but only one seller, there is no NE because the seller monopolizes the market and can keep raising the price. Refer to the appendix for detailed proofs.} in the market, the reselling game has a unique NE $\textbf{b}^*$. The NE $\textbf{b}^*$ can be achieved as the unique solution to the following optimization problem:
\begin{align*}
\text{P1: max}&\ \ \sum_{u_i}^{\mathcal{B}^t} \hat{U}_i^t(a_i^t) + \sum_{u_j}^{\mathcal{S}^t} \check{U}_j^t(a_j^t)\\
\text{subject to}&\ \ -q_i^t \le a_i^t,\ \  \sum_{u_i}^\mathcal{U}a_i^t=0.
\end{align*}
where approximate utility functions $\hat{U}_i^t(a_i^t)$ and $\check{U}_j^t(a_j^t)$ are:
\begin{equation}
\label{modified_utility}
\hat{U}_i^t(a_i^t) = (1-\frac{a_i^t}{\sum_{u_j\in \mathcal{S}^t}q_j^t})U_i^t(a_i^t) + \frac{1}{\sum_{u_j\in \mathcal{S}^t}q_j^t}\int_{0}^{f_i^ta_i^t}U_i^t(z)dz,
\end{equation}
\begin{equation}
\begin{aligned}
\check{U}_j^t(a_j^t) &= \frac{1}{\sum_{u_m\in \mathcal{S}^t,m \ne j}q_m^t}\int_{f_i^ta_j^t}^{0}U_j^t(z)dz\\ 
&- (1+\frac{a_j^t}{\sum_{u_m\in \mathcal{S}^t,m \ne j}q_m^t})U_j^t(a_j^t).
\label{modified_cost}
\end{aligned}
\end{equation}

\begin{proof}At the NE, no user can improve their payoff unilaterally, which means that $\forall u_i \in \mathcal{U}, \frac{\partial V_i^t}{\partial b_i^t} = 0$. 
For buyers, 
\begin{equation*}
\frac{\partial V_i^t}{\partial b_i^t} = \frac{\partial (U_i^t(a_i^t) - a_i^tp^t)}{\partial b_i^t}=0.
\end{equation*}
Thus,
\begin{align*}
\frac{\partial U_i^t(a_i^t)}{\partial b_i^t} &= 1\\
f_i^tU_i^{t'}(a_i^t) \cdot \frac{\partial a_i^t}{\partial b_i^t} &= 1\\
f_i^tU_i^{t'}(a_i^t) \cdot \frac{\sum_{k\ne i}^\mathcal{U}b_k^t\sum_j^{\mathcal{S}^t}q_j^t}{(\sum_k^\mathcal{U}b_k^t)^2} &= 1\\
f_i^tU_i^{t'}(a_i^t) \cdot (\frac{1}{p^t}-\frac{b_i^t}{p^t\sum_{u_j\in \mathcal{U}}b_j^t})&= 1.
\end{align*}
For sellers, a similar derivation applies. We have the first-order conditions for the NE as 
\begin{align}
\forall u_i \in \mathcal{B}^t,\ f_i^tU_i^{t'}(a_i^t)(1-\frac{a_i^t}{\sum_{u_j\in \mathcal{S}^t}q_j^t}) = p^t,\label{optimal_condition_buyer}\\
\forall u_j \in \mathcal{S}^t,\ f_j^tU_j^{t'}(a_j^t)(1+\frac{a_j^t}{\sum_{u_m\in \mathcal{S}^t,m \ne j}q_m^t}) = p^t. \label{optimal_condition_seller}
\end{align}
At the same time, from Eqs. (\ref{modified_utility}) and (\ref{modified_cost}) we have
\begin{equation*}
\forall u_i \in \mathcal{B}^t,\ \hat{U}_i^{t'}(a_i^t) = (1-\frac{a_i^t}{\sum_{u_j\in \mathcal{S}^t}q_j^t})f_i^tU_i^{t'}(a_i^t),
\end{equation*}
\begin{equation*}
\forall u_j \in \mathcal{S}^t,\ \check{U}_j^{t'}(a_j^t) = -(1+\frac{a_j^t}{\sum_{u_m\in \mathcal{S}^t,m \ne j}q_m^t})f_j^tU_j^{t'}(a_j^t).
\end{equation*}
Problem P1 is a strictly convex problem, where a unique optimal solution exists. Based on the KKT conditions, we have the optimality conditions of P1 as
\begin{equation}
\label{buyer_KKT}
\forall u_i \in \mathcal{B}^t,\ f_i^tU_i^{t'}(a_i^t)(1-\frac{a_i^t}{\sum_{u_j\in \mathcal{S}^t}q_j^t}) = \lambda,
\end{equation}
\begin{equation}
\label{seller_KKT}
\forall u_j \in \mathcal{S}^t,\ f_j^tU_j^{t'}(a_j^t)(1+\frac{a_j^t}{\sum_{u_m\in \mathcal{S}^t,m \ne j}q_m^t}) = \lambda,
\end{equation}
\begin{equation}
\label{total_KKT}
\sum_{u_i}^\mathcal{U}a_i^t=0.
\end{equation}

Thus, the first-order optimality conditions of P1 (Eq. (\ref{buyer_KKT})-(\ref{total_KKT})) are equivalent to the NE (Eq. (\ref{optimal_condition_buyer}) and (\ref{optimal_condition_seller})) when $p^t = \lambda$.
\end{proof}

\subsection{Iterative Bidding Algorithm to Reach the NE}

Based on Eq. (\ref{modified_utility}) and (\ref{modified_cost}), the modified payoff functions of buyers and sellers are respectively given as
\begin{equation}
\label{modified_buyer_payoff}
\hat{V}_i^t(b_i^t, p^t) = \hat{U}_i^t(b_i^t/p^t)-p^ta_i^t,
\end{equation}
\begin{equation}
\label{modified_seller_payoff}
\check{V}_j^t(b_j^t,p^t)=\check{U}_j^t(b_j^t/p^t) - p^ta_j^t.
\end{equation}
We now propose an iterative bidding algorithm to reach the NE, as presented in Alg. \ref{alg_IBA}. The algorithm focuses on the \textit{Choose Role} phase (lines 2-4) and the \textit{Bid} phase (lines 6-11). 

In the \textit{Choose Role} phase, each user $u_i$ reports their willingness to buy $w_i^t = {U_i^{t'}}(0)$ to the broker, where $w_i$ measures how fast the user's utility value grows with micro-purchase growth. The broker collects all the users' willingness and announces the average value $\overline{w^t}$ as a threshold to guide the users' choice that
\begin{equation}
u_i^{t}\in\left\{\begin{array}{ll}
\mathcal{S}^t, & \text { if } w_i^t < \overline{w^t},\\
\mathcal{B}^t, & \text { otherwise}.
\end{array}\right.
\label{role_criteria}
\end{equation}
After everyone has reported their role decisions, the \textit{Bid} phase starts once the broker determines that there is at least one buyer and two sellers.

Users make multiple rounds of bids based on market prices during the \textit{Bid} phase until the market price converges. Each round of the bidding has three stages: The broker announces a market price at stage 1; buyers and sellers bid at stage 2, according to the principle of maximizing their payoffs as in Eqs. (\ref{modified_buyer_payoff}) and (\ref{modified_seller_payoff}); the broker computes a new market price 
\begin{equation}
p^{t,[k+1]} = (p^{t,[k]}-\delta(\frac{\sum_{u_i}^{\mathcal{B}^t} b_i^{t,[k]} + \sum_{u_j}^{\mathcal{S}^t} b_j^{t,[k]}}{p^{t,[k]}}-\sum_{u_j\in \mathcal{S}^t}q_j^t))^+
\label{eq_price_update}
\end{equation}
based on the bids at stage 3.

\begin{algorithm}[t]
\label{alg_IBA}
\caption{Iterative Bidding Algorithm}
\For{each time slot}{
Users report their willingness to buy\;
Broker announces the average willingness\;
Users choose their role via (\ref{role_criteria}) and report\;
\If{there exist at least two sellers and one buyer}{
$k \leftarrow 0$;\ $p^{[k]} \leftarrow \text{initial\ price}$\;
\While{$k = 0\ ||\ \frac{|(p^{[k+1]}-p^{[k]})|}{p^{[k]}} > \epsilon$}{
\textbf{Stage 1}: Broker announces price $p^{[k]}$\;
\textbf{Stage 2}: Buyers obtain the optimal bids $b_i^{t,[k]}$ to maximize (\ref{modified_buyer_payoff}); Sellers obtain the optimal bids $b_j^{t,[k]}$ to maximize (\ref{modified_seller_payoff})\;
\textbf{Stage 3}: Upon the bids, broker updates the market price according to (\ref{eq_price_update})\;
}
Broker announces market-clearing price $p^{[k+1]}$\;
}\Else{Broker announces market closure\;}
}
\end{algorithm}

In stage 2, users bid for RB quotas for the next time slot to maximize their respective payoffs, in which their expected utility values are obtained based on the predicted future losses. More accurate predictions will lead users to make better bids, as the subsequent actual situation will be more in line with their expectations, and the resale mechanism will play a more positive role. To this end, we propose a heuristic function of the predicted loss as follows:
\begin{equation}
    pl_i^{t+1} = l_i^t - f_i^ta_i^t.
\label{eq_predict_1}
\end{equation}
Eq. (\ref{eq_predict_1}) specifies that if one does not participate in the resale, \textit{i.e.}, $a_i^t = 0$, then the immediate next-step loss will remain the same as the loss of the current time slot. In addition, we propose a future multi-step loss prediction as
\begin{equation}
    pl_i^{t+1} = l_i^t - (1 + \gamma + \gamma^2 + \gamma^3 + \cdots) * f_i^ta_i^t = l_i^t - \frac{1}{1 - \gamma} * f_i^ta_i^t,
\label{eq_predict_2}
\end{equation}
where a discount factor $\gamma \in (0, 1)$ indicates how that user considers future losses in relation to the current buffer situation.

\addtolength{\topmargin}{0.01in}

\section{Evaluation}
\label{chap_evaluation}

\subsection{Simulation Setup}

\begin{table}[b]
\vspace{-0.2cm}
\centering
\caption{Environmental Parameters}
\begin{tabular}{cl|cl}
\hline
Parameters & Values  & Parameters  & Values   \\ \hline
$H$        & 10 m    & $P^{tx}$    & 0.1 W    \\
$N$        & -96 dBm & $f_c$       & 2.4 GHz  \\
$W$        & 360 kHz & $c$         & 3.0 * 10\textasciicircum{}8 m/s \\ 
$T$        & 10 s    & $\epsilon$  & $10^{-5}$ \\ \hline
\end{tabular}
\label{parameters}
\end{table}

\begin{figure*}[tb]
    \centering
    \subfigure[Market price]{\includegraphics[scale=0.37]{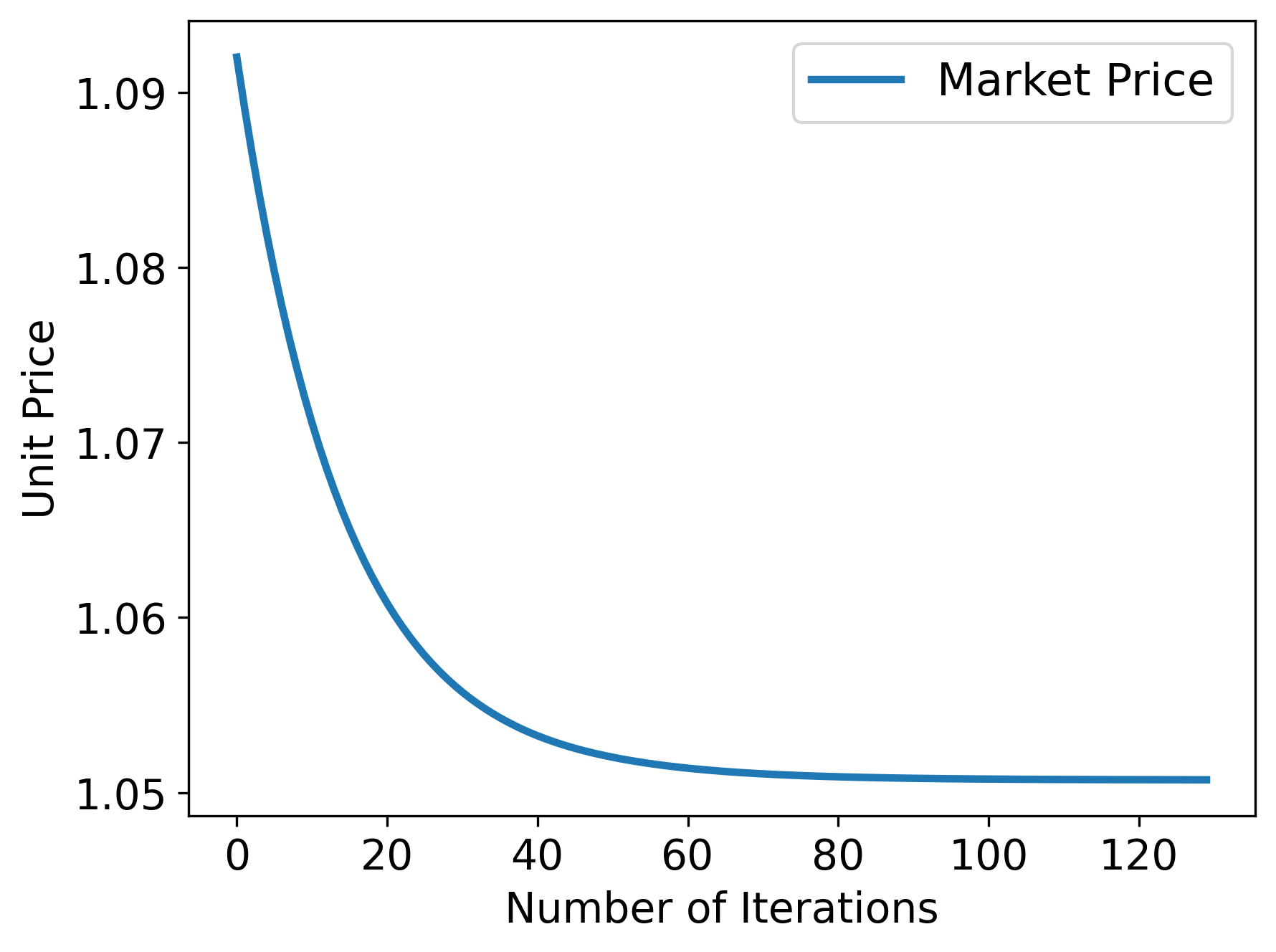}
    \label{price_conver}}
    \hspace{6pt}
    \subfigure[Requesting and sharing resources]{\includegraphics[scale=0.37]{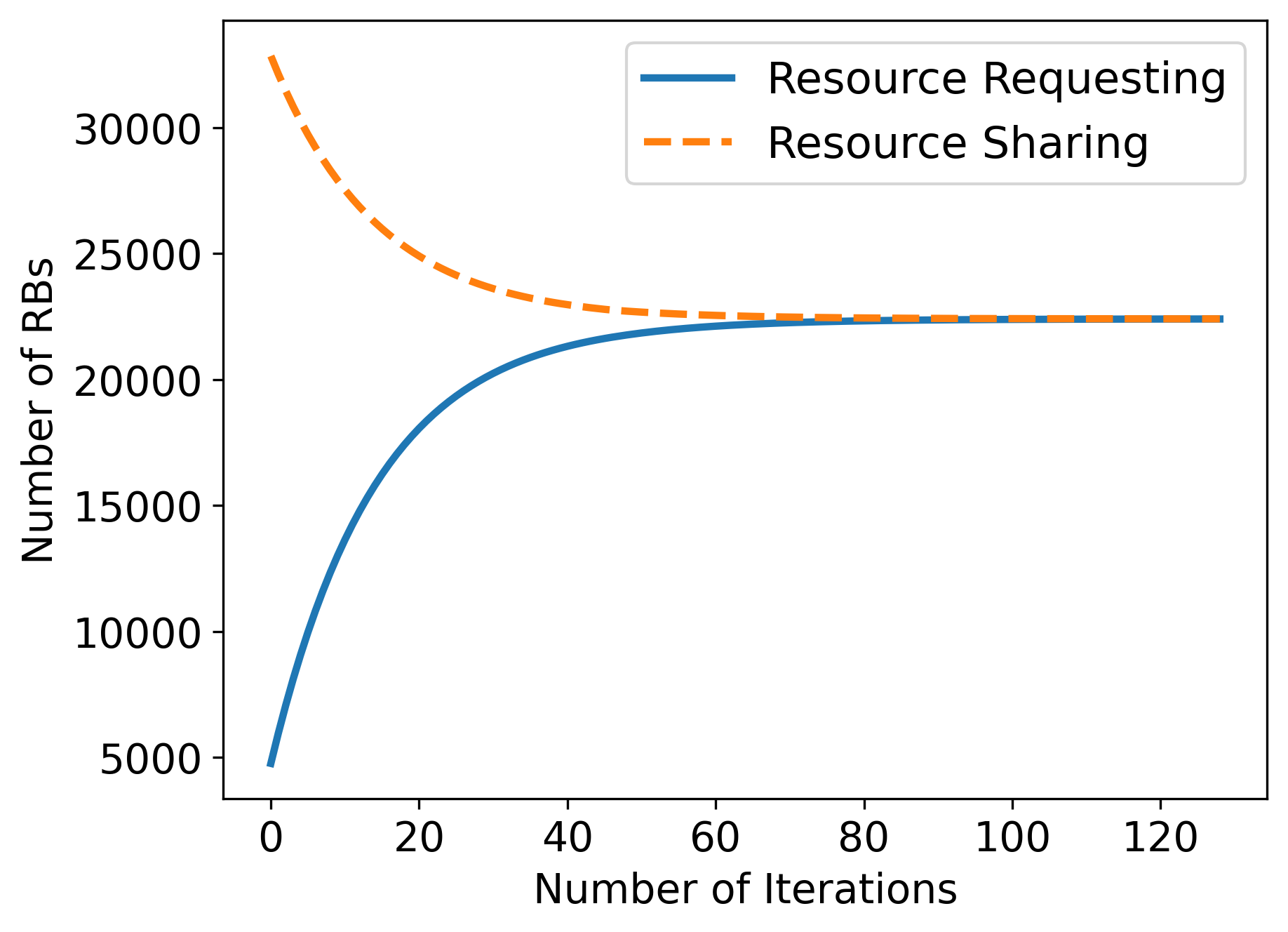}
    \label{res_conver}}
    \hspace{6pt}
    \subfigure[Social welfare]{\includegraphics[scale=0.37]{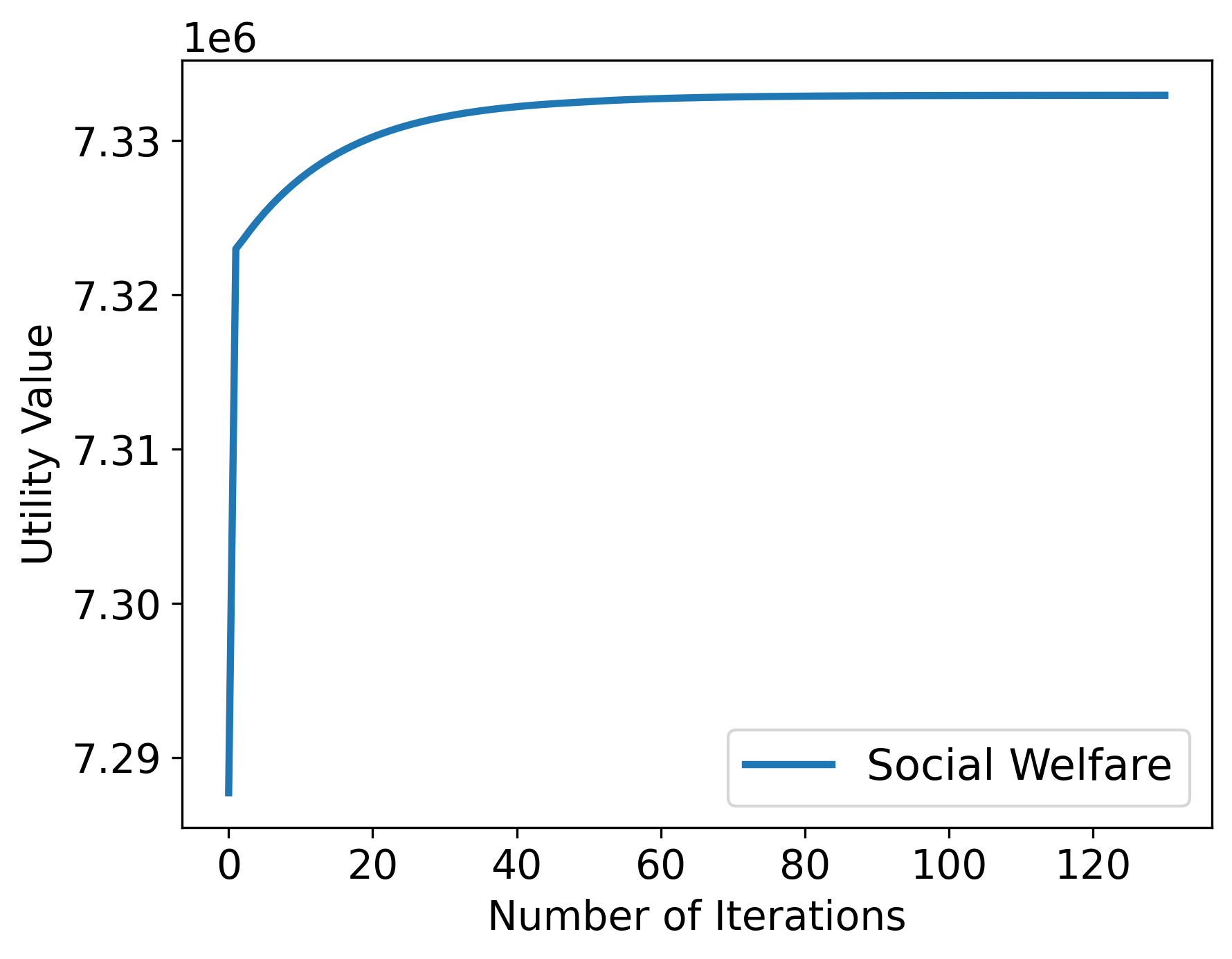}
    \label{wel_conver}}
    \caption{\textbf{Convergence process in iterative bidding}. A total of 130 rounds of bidding occurred throughout the process. The step size $\delta$ is set to $10^{-7}$, and the initial price is set to 1.095. During the first 20 rounds of bidding, the market price quickly reduced to 1.06 units. And over the next 110 rounds of bidding, the price falls more and more slowly, until it finally converges to a market-clearing price of about 1.05 units.}
\vspace{-0.2cm}
\end{figure*}

In this section, we evaluate our proposed reselling mechanism in a 100m $\times$ 100m square area, with the base station fixed at the center. Table \ref{parameters} shows the environmental parameters for the simulation. We assume 5 high-bandwidth (HB, e.g., similar to eMBB) users and 5 lower-rate (LR, e.g., similar to URLLC) users amongst 10 users, with utility functions
\begin{equation*}
    U_i^{t}(l_i^t)=s_i * (\sqrt{(D_{max} - l_i^t))} - \sqrt{D_{max}}),
\end{equation*}
where $s_i$ indicates how sensitive users are to the utility of loss relative to money. The $s_i$ of HB and LR users are uniformly generated in $[21, 23]$ and $[23, 25]$. According to the 3GPP technical specification \cite{3GPP_38.101-1}, the RB number of a 5 MHz channel bandwidth and 30 kHz subcarrier spacing\footnote{
An RB consists of 12 consecutive subcarriers.} is 11 per \textit{slot}\footnote{Distinct from the time slot defined in this paper.}, and there are 2000 \textit{slots} per second. So there are a total of $11 * 2000 * 10 = 220,000$ RBs per time slot. We assign each HB user 4,000 RBs, each LR user 40,000 RBs.

User mobility follows the random waypoint model \cite{johnson1996dynamic}, where users travel at a fixed speed of 10 m/slot in random directions. To capture the bursty nature of network data, we let user-generated data awaiting upload follow the Pareto distribution, which is a representative heavy-tailed distribution to model network traffic \cite{liu2022complexity}. The $d_i^t$ values for LR users vary from 10 to 100 Mb, with an average of 11 Mb, while the values for HB users vary from 100 to 150 Mb, with an average of 108 Mb. All the users have a buffer size of 1 Gb, and their initial empty buffer is randomly generated from 30 to 70 Mb. We conduct simulation experiments using the Python language\footnote{Code is available at https://github.com/RyderCRD/ORAN-Resale.}.

\subsection{Validity Verification}

We first experimentally verify the convergence of prices in the bidding process. Fig. \ref{price_conver} shows the price trend during the iterative bidding of a single time slot. Along with the convergence of market price, the total demand and total supply gradually become the same, as shown in Fig. \ref{res_conver}. Resource requests increase as the price decreases, because the cheaper the RBs, the more buyers want to buy. Conversely, the resource supply falls with the price, since the sellers want to sell less because of falling profits. In the final convergence, the total amount of buying and selling is equal, and the market clears.

Social welfare $\sum_{i=1}^N U_i(a_i^t)$ is obtained by summing the utility values of all the users, as the bids are zero-sum. Fig. \ref{wel_conver} shows its upward trend during the bidding process. The reselling mechanism allows relatively less needy users to sell their RB quotas to more needy users in exchange for monetary benefits, in which the difference in the degree of need is aligned. The buyers' utility values rise more than the sellers' utility values fall, which results in a rise in social welfare.

\subsection{Performance Analysis}

We compare our proposed reselling mechanism, integrating the iterative bidding algorithm with the heuristic role-decision scheme \textit{Heuristic} (loss prediction by Eq. (\ref{eq_predict_1})) and a variant \textit{Future} (by Eq. (\ref{eq_predict_2}), $\gamma = 0.9$) against two baselines: (1) \textit{Static}: the network slicing remains static, and there is neither trading nor RB re-allocation. (2) \textit{Random}: Users are randomly assigned as sellers or buyers for each round, who then bid \cite{tang2016double}. In our comparison, we evaluate performance by the following three key metrics: (1) data loss, (2) RB wastage, and (3) social welfare.

\begin{table}[b]
\vspace{-0.4cm}
    \caption{Numerical Results}
    \centering
    \scalebox{0.88}{
        \begin{tabular}{|c|cc|cc|c|}
        \hline
        \multirow{2}{*}{} & \multicolumn{2}{c|}{Loss}         & \multicolumn{2}{c|}{Wastage}        & \multirow{2}{*}{\begin{tabular}[c]{@{}c@{}}$\Delta$ Social\\ Welfare\end{tabular}} \\ \cline{2-5}
                          & \multicolumn{1}{c|}{No.} & Amount & \multicolumn{1}{c|}{No.} & Amount &                                                                              \\ \hline
        Static            & \multicolumn{1}{c|}{10235}  & 24.716 Gb      & \multicolumn{1}{c|}{1236}  & 14.487 Gb      & 0.0                                                                            \\ \hline
        Random            & \multicolumn{1}{c|}{5695}   & 63.186 Gb      & \multicolumn{1}{c|}{4618}  & 71.073 Gb      & $-4.353 * 10^{6}$                                                                            \\ \hline
        Heuristic         & \multicolumn{1}{c|}{2440}   & 17.213 Gb      & \multicolumn{1}{c|}{589}   & 7.1552 Gb      & $4.417 * 10^{6}$                                                                            \\ \hline
        Future            & \multicolumn{1}{c|}{2467}   & 17.186 Gb      & \multicolumn{1}{c|}{586}   & 7.1445 Gb      & $10.258 * 10^{6}$                                                                            \\ \hline
        \end{tabular}
    }
\label{table_num}
\end{table}

\begin{figure*}[tb]
    \centering
    \subfigure[Static]{\includegraphics[scale=0.291]{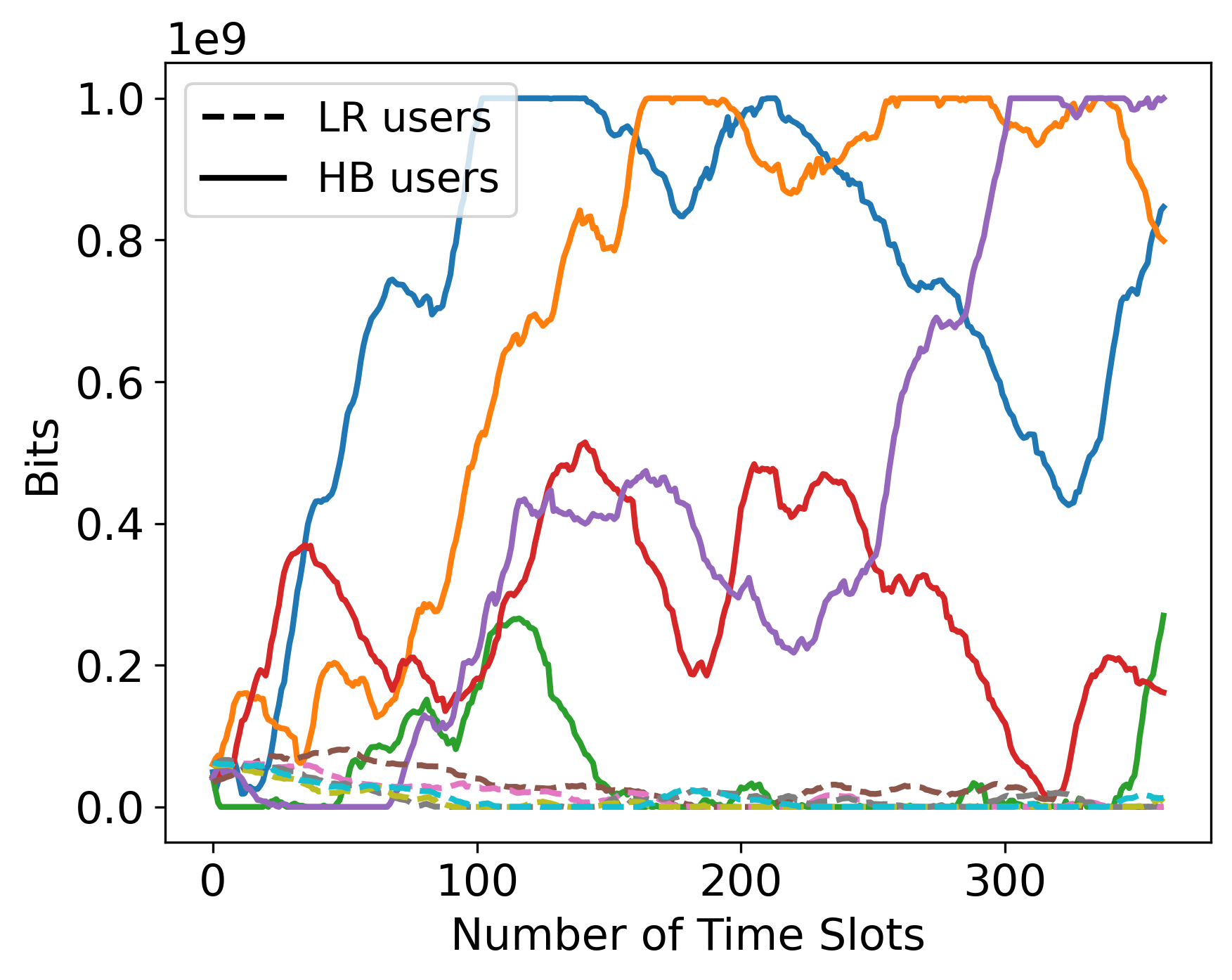}
    \label{buf_sta}}
    \subfigure[Random]{\includegraphics[scale=0.291]{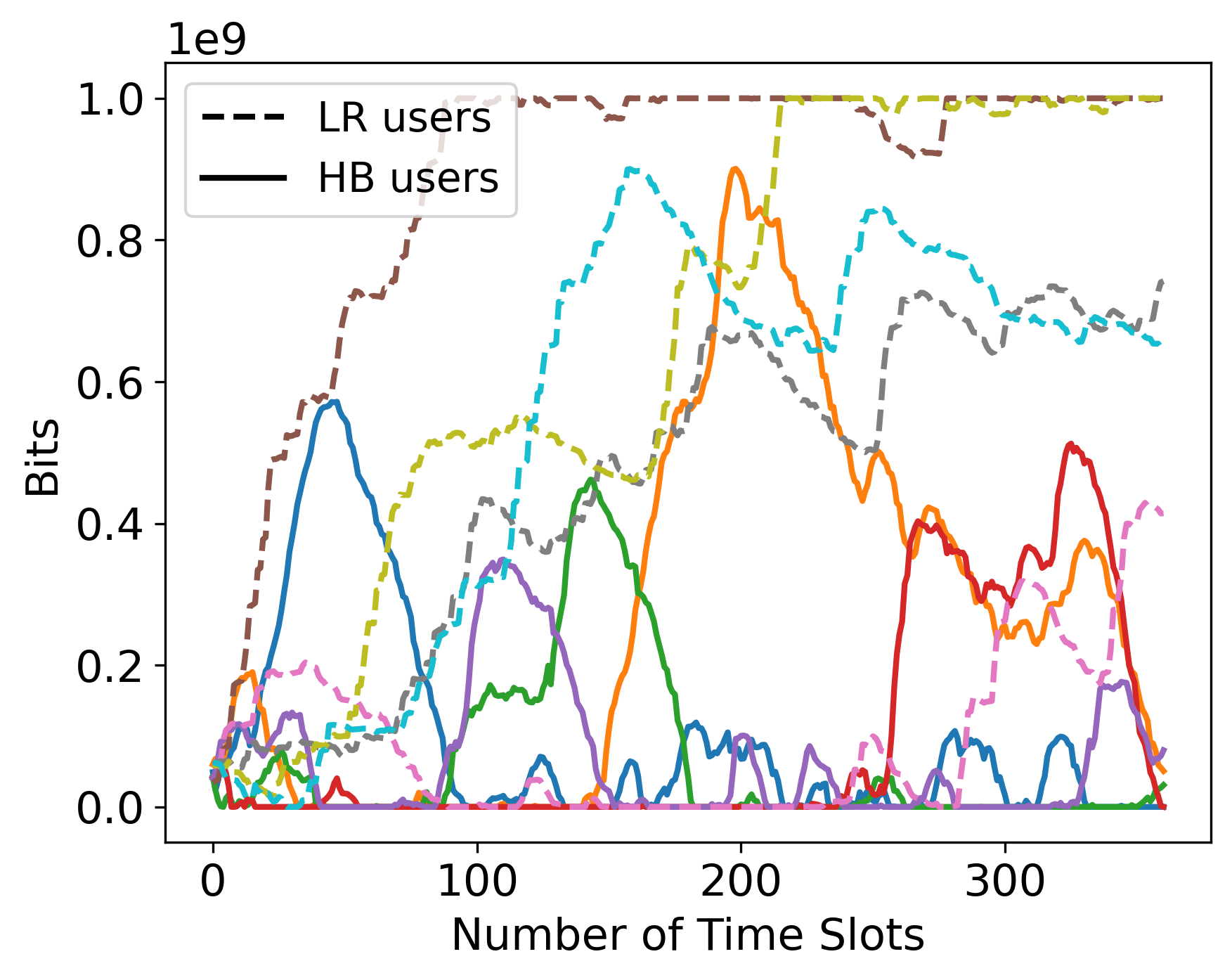}
    \label{buf_ran}}
    \subfigure[Heuristic]{\includegraphics[scale=0.291]{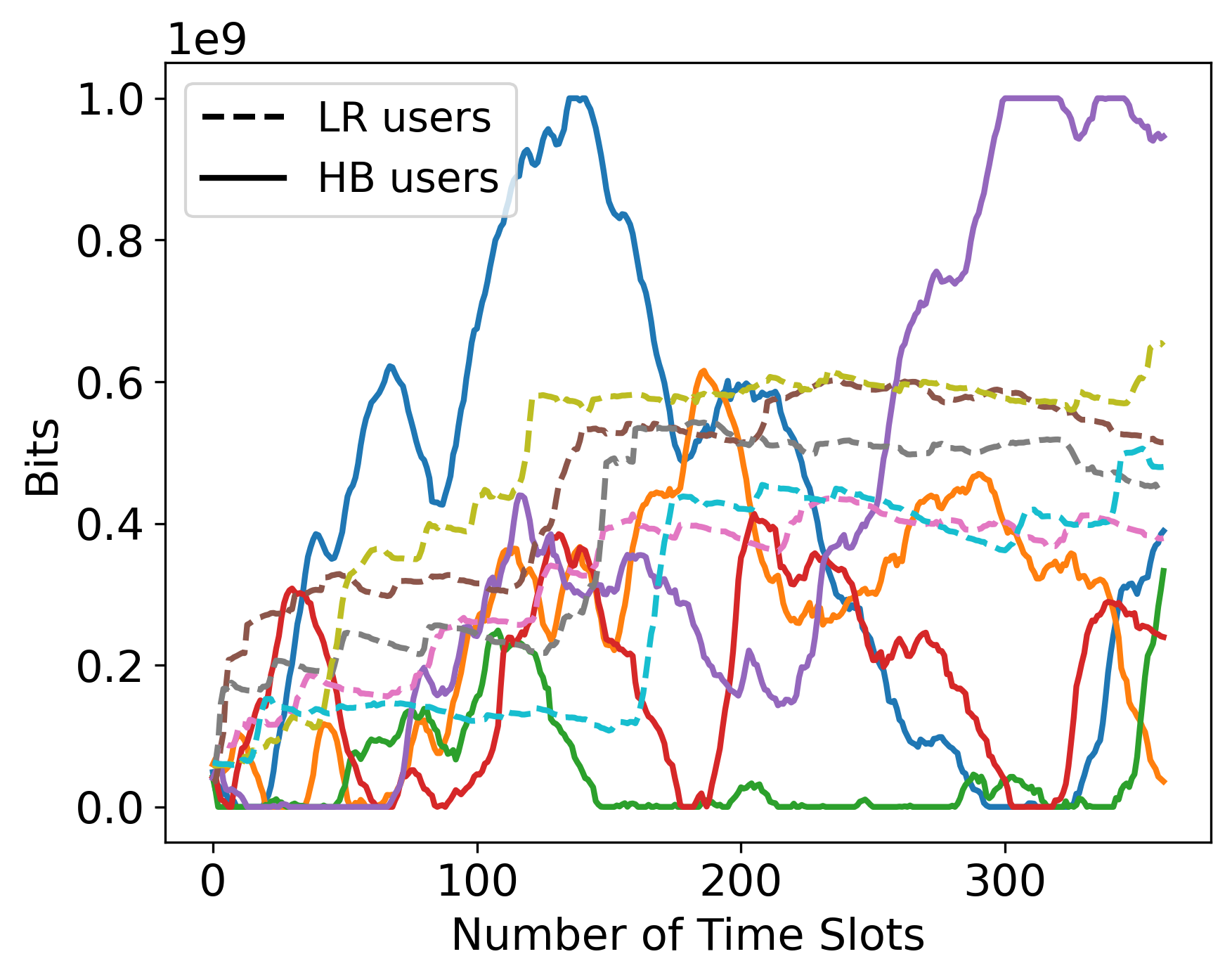}
    \label{buf_heu}}
    \subfigure[Future]{\includegraphics[scale=0.291]{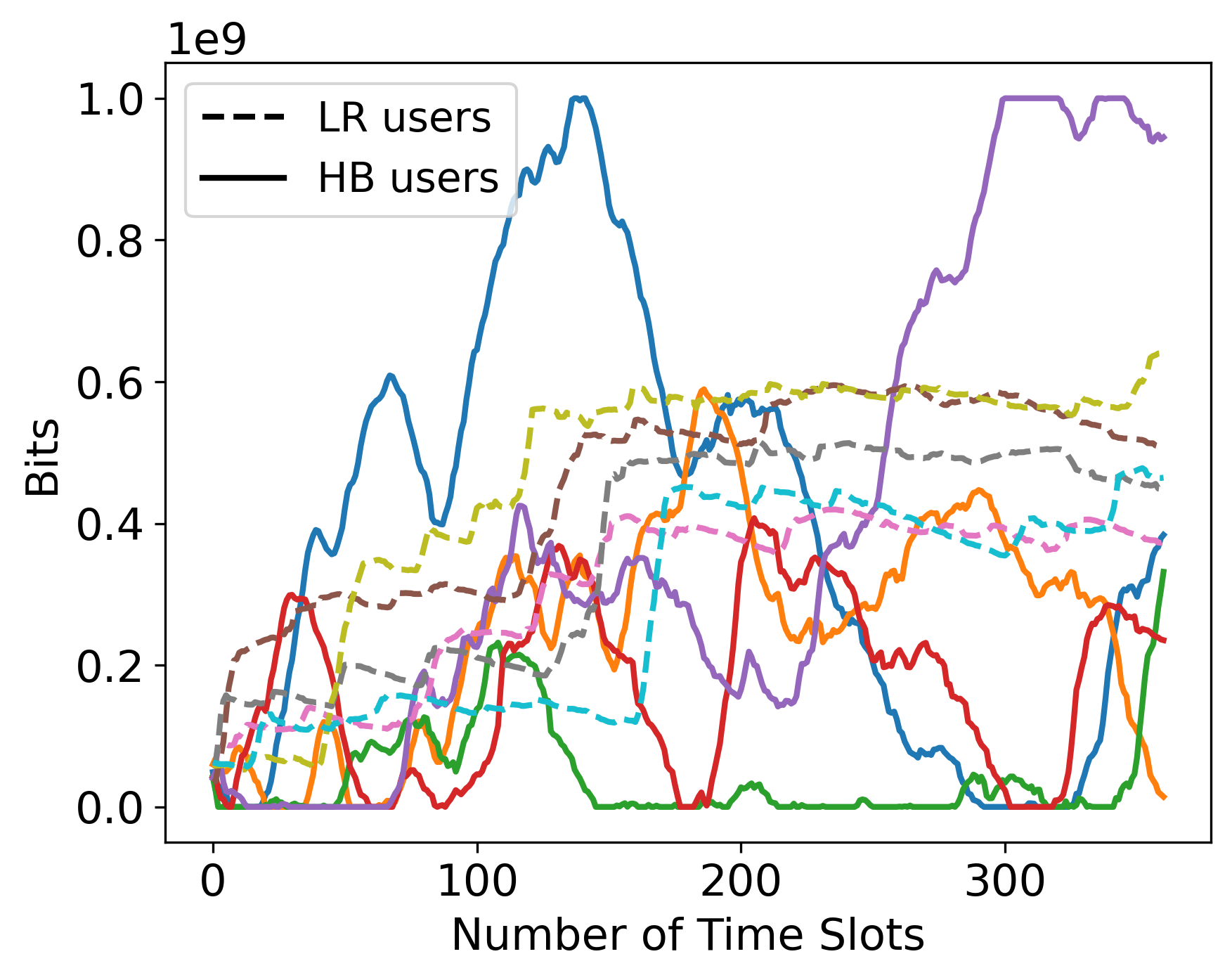}
    \label{buf_fut}}
    \subfigure[Static]{\includegraphics[scale=0.29]{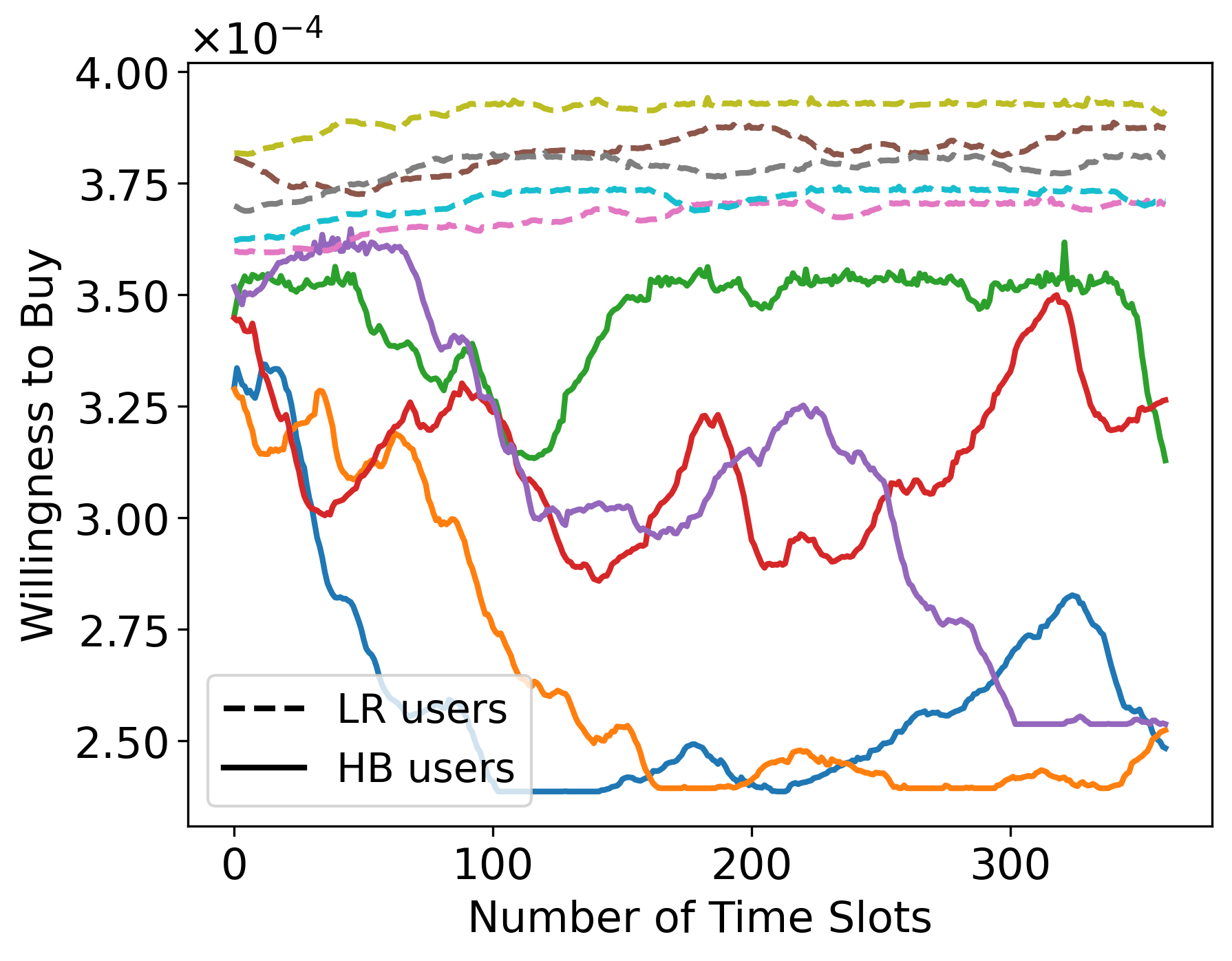}
    \label{wil_sta}}
    \subfigure[Random]{\includegraphics[scale=0.29]{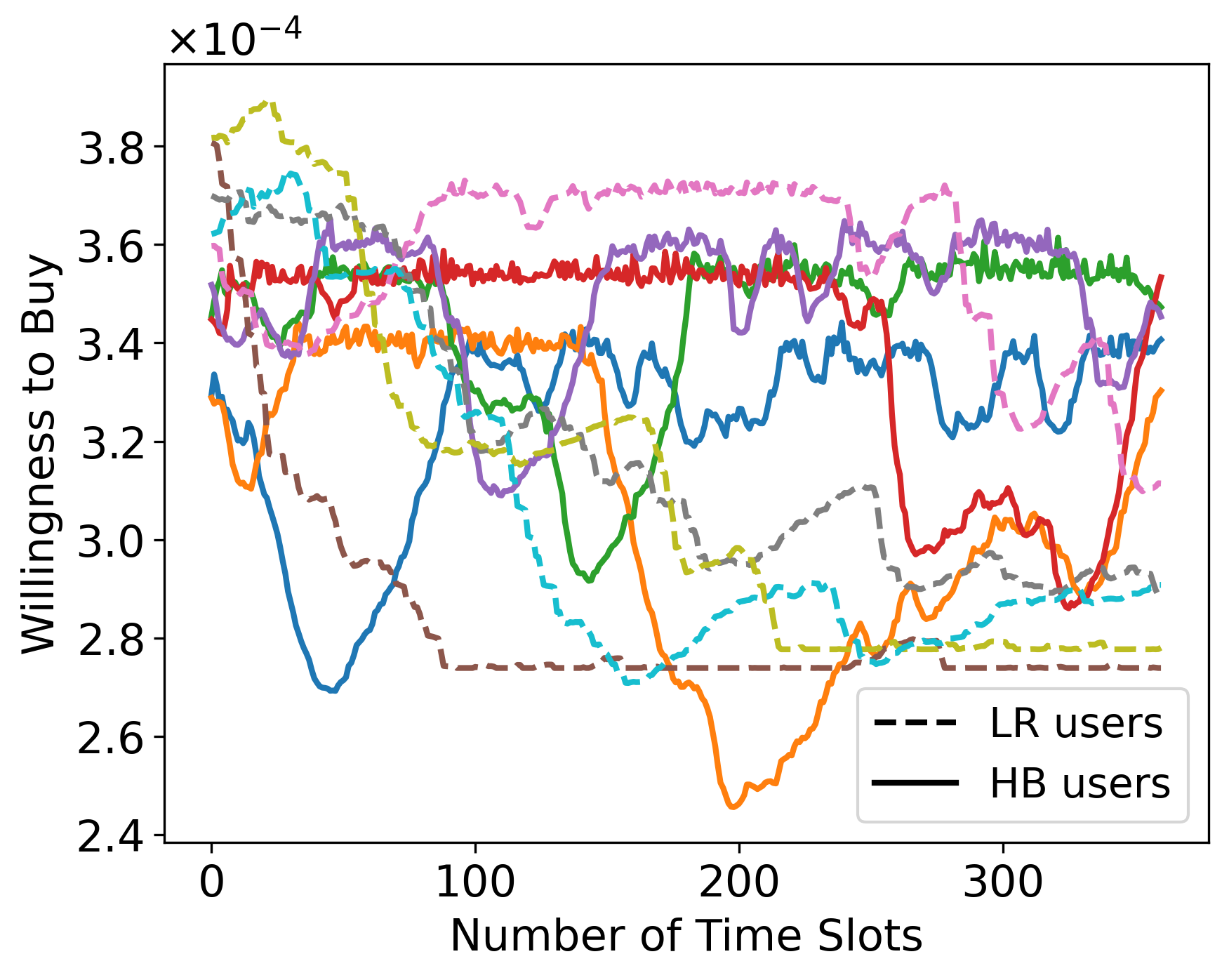}
    \label{wil_ran}}
    \subfigure[Heuristic]{\includegraphics[scale=0.29]{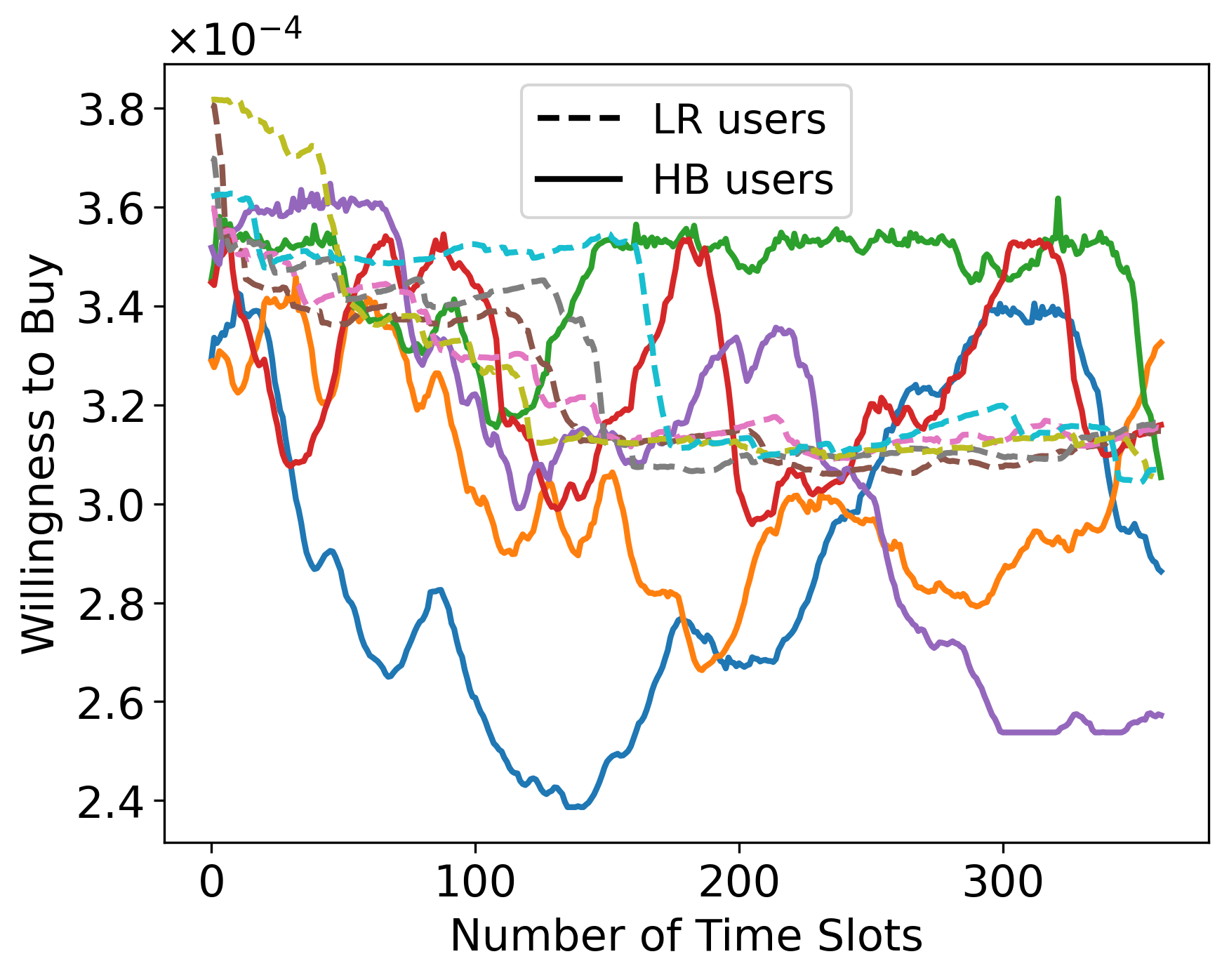}
    \label{wil_heu}}
    \subfigure[Future]{\includegraphics[scale=0.29]{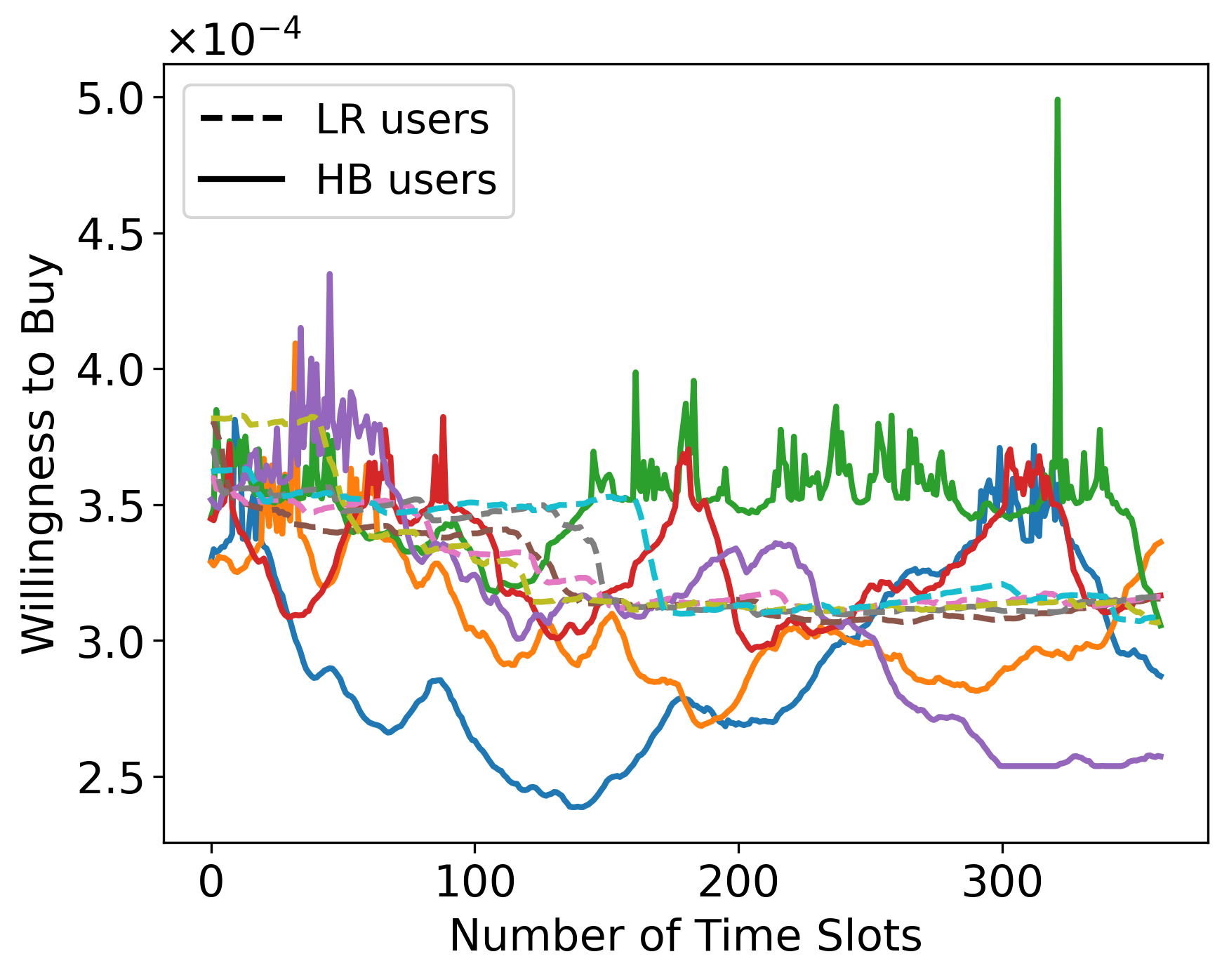}
    \label{wil_fut}}
    \caption{\textbf{Empty buffer and willingness trends}. Lines in different colors correspond to the trends of different users. The willingness to buy value shows a high correlation with the user's current buffer availability. The emptier the user's buffer, the lower the willingness to buy. When the buffer is empty, the user's willingness to buy is minimized. When the buffer is full, the user's willingness to buy is high and could be even higher due to data loss.}
    \label{fig_trends}
\vspace{-0.3cm}
\end{figure*}

Table \ref{table_num} shows the numerical results of the simulation for a total of 4320 time slots in 12 hours. To accelerate convergence, the step size is set to $10^{-6}$. The Static scheme exhibits the highest number of losses (10,235) but a relatively low loss amount (24.716 Gb), suggesting frequent but small-scale losses. In contrast, the Random scheme shows fewer losses (5,695) but a significantly higher loss amount (63.186 Gb), indicating less frequent but more severe losses. The Heuristic and Future schemes perform similarly, with the lowest numbers of losses (2,440 and 2,467, respectively) and loss amounts (around 17.2 Gb), demonstrating their effectiveness in minimizing both the frequency and the scale of the losses. For wastage, the Static scheme achieves a moderate number (1,236) and amount (14.487 Gb), while the Random scheme performs poorly with high wastage numbers (4,618) and amounts (71.073 Gb). The Heuristic and Future schemes again excel, with the lowest wastage numbers (589 and 586) and amounts (around 7.15 Gb), highlighting their efficiency in reducing waste.

The social welfare change reveals stark differences among the schemes. The Static scheme serves as the baseline having no change (0.0). The Random scheme results in a significant decline, indicating a detrimental impact on the social welfare. In contrast, the Heuristic scheme improves the social welfare, although the Future scheme outperforms it by a substantial margin. This suggests that while the Heuristic scheme is effective, the Future scheme offers even greater benefits, likely due to its forward-looking characteristic. The superior social welfare outcomes of the Heuristic and Future schemes align with their better performance in minimizing losses and wastage, underscoring their overall advantages.

In order to compare the schemes in more detail, we track the empty buffer $e_i^t$ and willingness-to-buy $w_i^t$ in the 360 slots of the first hour. Fig. \ref{fig_trends} (a), (b), (c), and (d) show the user's buffer changes under the Static, Random, Heuristic, and Future schemes, respectively. Data loss occurs when there is no empty buffer, while RB wastage occurs when the buffer is fully empty.

There is a significant difference in empty buffer trends between LR and HB users in the Static scheme. The empty buffer for LR users remains very low, never exceeding 10\% of the total buffer, while the empty buffer for HB users reaches 100\% sometimes. This implies resources are not well utilized, that the HB users generate RB wastage while the LR users are unable to accommodate the continuous influx of data with their limited buffer, resulting in overflows and data loss. The opposite situation occurs in the Random scheme, where the LR users waste RBs, and HB users suffer data loss. Compared to the two baselines, the Heuristic and Future schemes show better buffer management. Most users (especially LR users who pursue reliability) are able to keep the amount of empty buffers within reasonable bounds, thus avoiding data losses and RB wastage.

The trends of willingness to buy under the different schemes are illustrated in Fig. \ref{fig_trends} (e), (f), (g), and (h), respectively. In the Static scheme, the willingness to buy is consistently high among all the LR users who lack RB quotas but are unable to buy more. With reselling enabled, the trends of willingness to buy in the other schemes are more dynamic. Users are more sensitive to losses in the Future scheme compared to the Heuristic scheme, and the rise in users' willingness to buy when losses occur is more pronounced. Such sensitivity helps the users make more timely buffer adjustments.

\section{Conclusion}
\label{chap_conclusion}

This paper presents a game-theoretic 5G resource reselling mechanism where users dynamically trade RB quotas based on buffer status and predicted losses, improving social welfare despite selfish behavior. We prove the existence and uniqueness of the NE and introduce a heuristic role-selection strategy with two predictive functions: short-term loss estimation and multi-step future loss forecasting. Experiments show that our approach reduces loss by 30.5\% and wastage by 50.7\% while significantly increasing social welfare, compared to the static baseline.

Three key directions warrant further investigation: (1) incorporating trajectory and traffic pattern priors to enhance loss prediction accuracy and bidding performance, (2) extending the framework to multi-BS scenarios with handover support for broader applicability, and (3) generalizing the current uplink-focused mechanism to include downlink resource trading.  These extensions would further enhance the practicality.

\appendix
\subsection{Lemma 1 and the Proof}

\noindent \textit{Lemma 1}: When there is only one seller, i.e., $|S^t|=1$, there is no Nash equilibrium.

\begin{proof}
We prove by contradiction. Suppose there is a single seller $u_i$, then according to Eq. (\ref{seller_bid}), for sellers we have
\begin{equation}
a_i^t = \frac{b_i^t}{p^t} - q_i^t.
\label{lemma_seller_amount}
\end{equation}
At the market-clearing point, total supply equals total demand, and we have $\sum_{u_j}^{\mathcal{B}^t} a_j^t = -\sum_{u_k}^{\mathcal{S}^t} a_k^t$. Hence,
\begin{equation*}
\sum_{u_j}^\mathcal{U}\frac{b_j^t}{p^t}=\sum_{u_k}^{\mathcal{S}^t}q_k^t.
\end{equation*}
\begin{equation}
p_t = \frac{\sum_{u_j}^\mathcal{U}b_j^t}{\sum_{u_k}^{\mathcal{S}^t}q_k^t}.
\label{lemma_price}
\end{equation}
Substituting Eq. (\ref{lemma_price}) into Eq. (\ref{lemma_seller_amount}) yields
\begin{equation}
a_i^t = \frac{b_i^t\sum_{u_k}^{\mathcal{S}^t}q_k^t}{\sum_{u_j}^\mathcal{U}b_j^t} - q_i^t.
\label{lemma_substituted_amount}
\end{equation}
Since there's only one seller, we can rewrite Eq. (\ref{lemma_substituted_amount}) (using $\sum_{u_k}^{\mathcal{S}^t}q_k^t = q_i^t$ and $\sum_{u_k}^{\mathcal{S}^t}b_k^t = b_i^t$), as
\begin{equation*}
a_i^t = \frac{b_i^tq_i^t}{\sum_{u_j}^{\mathcal{B}^t}b_j^t + b_i^t} - q_i^t = \frac{-q_i^t\sum_{u_j}^{\mathcal{B}^t}b_j^t}{\sum_{u_j}^{\mathcal{B}^t}b_j^t + b_i^t}.
\end{equation*}
Thus, we have $u_i$'s payoff as $V_i^t(b_i^t,\textbf{b}_{-i}^t)$
\begin{align*}
&=U_i^t(\frac{-q_i^t\sum_{u_j}^{\mathcal{B}^t}b_j^t}{\sum_{u_j}^{\mathcal{B}^t}b_j^t+b_i^t})-(\frac{\sum_{u_j}^{\mathcal{B}^t}b_j^t + b_i^t}{q_i^t}*\frac{-q_i^t\sum_{u_j}^{\mathcal{B}^t}b_j^t}{\sum_{u_j}^{\mathcal{B}^t}b_j^t + b_i^t}),\\
&=U_i^t(\frac{-q_i^t\sum_{u_j}^{\mathcal{B}^t}b_j^t}{\sum_{u_j}^{\mathcal{B}^t}b_j^t+b_i^t})+\sum_{u_j}^{\mathcal{B}^t}b_j^t.
\end{align*}
Therefore,
\begin{equation*}
\frac{\partial V_i^t(b_i^t,\textbf{b}_{-i}^t)}{\partial b_i^t} > 0,
\end{equation*}
meaning that the seller's payoff is strictly increasing with respect to $b_i^t$. The seller $u_i$ can raise the magnitude of the bid infinitesimally to boost the payoff value, i.e, the seller can profitably deviate. This shows that if there is only one seller, there is no Nash equilibrium.
\end{proof}

\bibliographystyle{IEEEtran}
\bibliography{GLOBECOM_REF}

\end{document}